\documentclass[aps,pra,superscriptaddress]{revtex4}
\usepackage{amsmath,epsfig}
\usepackage{graphicx}

\def\be{\begin{equation}}
\def\ee{\end{equation}}
\def\bea{\begin{eqnarray}}
\def\eea{\end{eqnarray}}
\newcommand{\ket}[1]{\mbox{$|#1\rangle$}}
\newcommand{\bra}[1]{\mbox{$\langle#1|$}}

\def\bfr{{\bf r}}
\def\bfd{{\bf d}}

\def\bfE{{\bf E}}

\def\bfp0{{\bf{p_0}}}

\def\Gammatotal{\gamma_{\footnotesize\textrm{total}}}
\newcommand{\opdagger}[2]{\mbox{$\hat{#1}_{#2}^{\dagger}$}}
\newcommand{\op}[2]{\mbox{$\hat{#1}_{#2}$}}

\begin{document}
\title{Broad-band spectral control of single photon sources using a nonlinear
photonic crystal cavity}

\author{Murray W.\ McCutcheon$^*$}
\affiliation{School of Engineering and Applied Sciences, Harvard University,
Cambridge, MA 02138}
\thanks{These authors contributed equally to this work.}
\email{murray@seas.harvard.edu}

\author{Darrick E.\ Chang$^*$}
\affiliation{Institute for Quantum Information and Center for the
Physics of Information, California Institute of Technology,
Pasadena, CA 91125} 

\author{Yinan Zhang}
\affiliation{School of Engineering and Applied Sciences, Harvard University,
Cambridge, MA 02138}

\author{Mikhail D.\ Lukin}
\affiliation{Physics Department, Harvard University, Cambridge, MA
02138}

\author{Marko Lon\v{c}ar}
\affiliation{School of Engineering and Applied Sciences, Harvard University,
Cambridge, MA 02138}

\date{\today}

\begin{abstract}
Motivated by developments in quantum information science, much
recent effort has been directed toward coupling individual quantum
emitters to optical microcavities. Such systems can be used to
produce single photons on demand, enable nonlinear optical
switching at a single photon level, and implement functional nodes
of a quantum network, where the emitters serve as processing nodes
and  photons are used for long-distance quantum communication. For
many of these practical applications, it is important to develop
techniques that allow one to generate outgoing single photons of
desired frequency and bandwidth, enabling hybrid networks
connecting different types of emitters and long-distance
transmission over telecommunications wavelengths. Here, we propose
a novel approach that makes use of a nonlinear optical resonator,
in which the single photon originating from the atom-like emitter
is directly converted into a photon with desired frequency and
bandwidth using the intracavity nonlinearity. As specific
examples, we discuss a high-finesse, TE-TM double-mode photonic
crystal cavity design that allows for direct generation of single
photons at telecom wavelengths starting from an InAs/GaAs quantum
dot with a $950$~nm transition wavelength, and a scheme for direct
optical coupling of such a quantum dot with a diamond
nitrogen-vacancy center at $637$~nm.
\end{abstract}

\maketitle
\nopagebreak

Nonlinear optical frequency conversion is widely used in fields as
diverse as ultrahigh-resolution imaging~\cite{campagnola99} and
telecommunications, as it allows for the generation of light in
parts of the spectrum for which there are no convenient sources.
Common implementations include optical parametric oscillators to
make tunable femtosecond lasers in the infrared, and conversion of
the 1064 nm Nd:YAG laser to make green laser sources via second-harmonic 
generation. Recently, nonlinear photonic crystal
cavities~\cite{McCutcheon_PRB, rodriguez07,bravo-abad07,
Liscidini06,Cowan05} have emerged as promising systems in which
similar nonlinear functionalities can be achieved at micron
scales, which would enable the miniaturization of optical devices
onto integrated platforms. While the majority of such work focuses
on conversion of classical fields, these systems are now being
applied to quantum optics and quantum information
science~\cite{vandevender04,langrock05,tanzilli05,vandevender07}.

In recent years, there has also been a concerted research effort
to develop on-demand single-photon sources using single quantum
emitters strongly coupled to resonant optical
microcavities~(cavity QED)~\cite{michler00,pelton02,mckeever04}.
The strong coupling of the emitter to a resonant cavity
results in preferential emission into the cavity mode of a single
photon with frequency near the atomic resonance. Connecting pairs
of such systems would form the basis for distributed quantum
networks, where the emitters serve as processors and photons carry
information between the nodes~\cite{Cirac}. In practice, however,
the photon emission occurs at wavelengths determined by the atomic
resonance frequency. This is impractical, as it does not exploit
the low-loss telecom frequency band for long-distance transmission
and requires all emitters in a quantum network to be identical.

Here, we describe a novel approach to generate single photons with
controllable wavelength and bandwidth. Our approach makes use of
an integrated nonlinear optical cavity in which optical emission
is directly frequency-shifted into the desired domain using
intracavity nonlinear optical processes. This cavity-based
generation technique is quite robust in that the maximum
efficiency does not depend on an explicit phase-matching
condition~\cite{boyd92}, as would occur in an extended nonlinear
crystal or fiber, but rather only on the ratio of the cavity
quality factor to mode volume~($Q/V$).
As an example, we demonstrate a novel double mode TE-TM cavity
design in a GaAs photonic crystal that is well-suited for the
conversion of photons from quantum dots to the telecom band.  We also
present a similar GaP-based design for
direct coupling between a nitrogen-vacancy center in
diamond~\cite{Dutt,Santori,Gaebel06,Hanson} and an InAs/GaAs
quantum dot~\cite{Srinivasan07, Hennessy07, Englund07}, which
could enable practical realization of a heterogeneous quantum
network. In addition to effective wavelength control of single
photons~\cite{vandevender04,langrock05,tanzilli05,vandevender07},
our approach enables the manipulation of their bandwidth, which is
important for fast communication.

\section*{The concept of single-photon spectral control}

\begin{figure*}[b]
\begin{center}
\includegraphics[width=13cm]{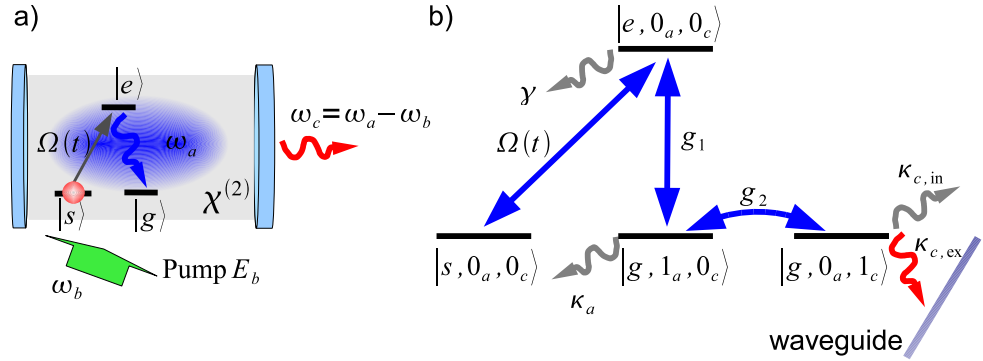}
\end{center}
\caption{\textbf{Schematic of single-photon frequency conversion}
a) A single three-level emitter is coupled to a double-mode cavity
that possesses a $\chi^{(2)}$ nonlinearity.  After excitation, the
emitter emits a photon into the cavity at frequency $\omega_a$.  When the
cavity is irradiated by the pump beam at $\omega_b$, the photon is converted
to a second cavity mode at frequency $\omega_c$.
b)  Level diagram: coherent coupling
strengths are indicated with blue arrows, while gray arrows denote
undesirable loss mechanisms. The emitter is controllably pumped
from initial state $\ket{s}$ via an external laser field
$\Omega(t)$ to excited state $\ket{e}$. The excited state
$\ket{e}$ can reversibly emit a single photon into cavity mode
$a$~(while bringing the atom into state $\ket{g}$) at a rate
$g_1$, and can also decay into free space at rate $\gamma$. Mode
$a$ has an inherent decay rate given by $\kappa_a$. The
nonlinearity allows the photon in mode $a$ to be converted to one
in mode $c$ at a rate $g_2$ when the cavity is pumped by a laser
of frequency $\omega_{b}=\omega_{a}-\omega_{c}$. The leakage rate
of the frequency-converted photon at $\omega_c$ is split up into
undesirable channels~($\kappa_{c,in}$) and desirable out-coupling
to a nearby waveguide~($\kappa_{c,ex}$).\label{fig:levels}}
\end{figure*}

We first discuss the general protocol for generating single
photons on demand at arbitrary frequencies using a nonlinear
double-mode cavity, and introduce a simple theoretical model to
derive the efficiency of the process. The system of interest is
illustrated schematically in Fig.~\ref{fig:levels}. As in standard
cavity-based single-photon generation
protocols~\cite{Cirac,mckeever04}, a single three-level atom~(or
any other quantum dipole emitter) is resonantly coupled to one
mode~(here denoted $a$) of an optical cavity. The emitter is
initialized in metastable state $\ket{s}$, and an external laser
field with controllable Rabi frequency $\Omega(t)$ couples
$\ket{s}$ to excited state $\ket{e}$. The transition between
$\ket{e}$ and ground state $\ket{g}$ is resonantly coupled to
cavity mode $a$~(frequency $\omega_a$), with a single-photon Rabi
frequency $g_1$. The relevant decay mechanisms~(illustrated with
gray arrows in the figure) are a leakage rate $\kappa_a$ for
photons to leave cavity mode $a$, and a rate $\gamma$ that
$\ket{e}$ spontaneously emits into free space rather than into the
cavity. Conventionally, in absence of an optical nonlinearity, the
control field $\Omega(t)$ creates a single atomic excitation at
some desired time in the system, which via the coupling $g_1$ is
converted into a single, resonant cavity photon. This photon
eventually leaks out of the cavity and constitutes an outgoing,
resonant single photon generated on demand whose spatial
wave-packet can be shaped by properly choosing
$\Omega(t)$~\cite{Cirac}.

In our system, the cavity is also assumed to possess a second mode
$c$ with frequency $\omega_c$, and our goal is to induce the
single photon to exit at this frequency rather than $\omega_a$.
This can be achieved, provided that the cavity medium itself
possesses a second-order~($\chi^{(2)}$) nonlinear susceptibility,
by applying a classical pump field to the system at the difference
frequency $\omega_{b}=\omega_{a}-\omega_{c}$. The induced coherent
coupling rate between modes $a$ and $c$ is denoted $g_2$. The
field $b$ need not correspond to a cavity mode. Mode $c$ has a
photon leakage rate, which we separate into an ``inherent'' rate,
$\kappa_{c,in}$, and a ``desirable'' (extrinsic) rate,
$\kappa_{c,ex}$. $\kappa_{c,in}$ characterizes the natural leakage
into radiation modes and also absorption losses, and can be
expressed in terms of the (unloaded) cavity quality factor as
$\kappa_{c,in} = \omega_c/2Q_{c}$. $\kappa_{c,ex}$ characterizes
the out-coupling rate into any external waveguide used for photon
extraction. The total leakage of mode $c$ is then
$\kappa_{c}=\kappa_{c,in}+\kappa_{c,ex}$.

More quantitatively, the effective Hamiltonian for the system~(in
a rotating frame) is given by
\bea H_I & = & H_{c}+H_{loss},\nonumber \\ H_{c} & = & \hbar g_{1}
(\sigma_{eg}a_{a}+\sigma_{ge}a^{\dagger}_{a})+\hbar\Omega(t)(\sigma_{es}+\sigma_{se})+\hbar
g_{2}(a^{\dagger}_{a}a_{c}+a_{a}a^{\dagger}_c),\nonumber \\
H_{loss} & = &
-\frac{i\gamma}{2}\sigma_{ee}-\frac{i\kappa_a}{2}a^{\dagger}_{a}a_{a}-\frac{i(\kappa_{c,ex}+\kappa_{c,in})}{2}a^{\dagger}_{c}a_{c},
\label{eq:HI} \eea
where $H_c$ describes the coherent part of the system
evolution~(for simplicity we take $g_{1,2},\Omega$ to be real),
and $H_{loss}$ is a non-Hermitian term characterizing the losses.
$\sigma_{ij}=\ket{i}\bra{j}$ are atomic operators, while $a_{i}$
is the photon annihilation operator for mode $i$. The vacuum Rabi
splitting $g_1$ can be written in the form
$g_1=\bfd\cdot\bfE_{a}(\bfr)/\hbar$, where $\bfd$ is the dipole
matrix element of the $\ket{g}$-$\ket{e}$ transition, and
$\bfE_{a}(\bfr)$ is the electric field amplitude per photon at the
emitter position $\bfr$. The electric field per photon in mode
$i=a,c$ is determined by the normalization
\be \frac{\hbar\omega_i}{2}=\int\,d\bfr
\epsilon_{0}\epsilon(\bfr)|\bfE_{i}(\bfr)|^2,\label{eq:normalization}
\ee
where $\epsilon(\bfr)$ is the dimensionless electric permittivity
of the material. The nonlinearity parameter is given
by~\cite{rodriguez07}
\be
g_{2}=-\frac{\epsilon_0}{\hbar}\int\,d\bfr\,\chi_{ijk}^{(2)}E_{a,i}^{\ast}\left(E_{b,j}E_{c,k}+E_{c,j}E_{b,k}\right).\label{eq:g2}
\ee
The amplitudes $E_{a,c}$ appearing above are normalized by
Eq.~(\ref{eq:normalization}), while $E_{b}$ is the classical pump
amplitude. Importantly, one can compensate for a small nonlinear
susceptibility $\chi^{(2)}$ or field overlap~(phase matching)
simply by using larger pump amplitudes $E_b$ to achieve a desired
$g_2$ strength.

For a system initialized in $\ket{s}$, there can never be more
than one excitation, and the system generally exists as a
superposition of having the system in state $\ket{s}$ or
$\ket{e}$~(with no photons) or having a photon in one of the modes
$a,c$~(and the emitter in $\ket{g}$),
\be
\ket{\psi(t)}=c_{s}(t)\ket{s}+c_{e}(t)\ket{e}+c_{a}(t)\ket{1_a}+c_{c}(t)\ket{1_c}.\label{eq:wavefn}
\ee
The system is initialized to $c_{s}(0)=1$ with all other
$c_{i}(0)=0$ and the time evolution is given by
$\dot{c}_{j}=-(i/\hbar)\bra{j}H_{I}\ket{\psi(t)}$. In this
effective wave-function approach, provided that $\ket{s}$ is
always driven, $\sum_{j}|c_{j}|^{2}{\rightarrow}0$ as
$t{\rightarrow}\infty$ due to losses, which can be connected with
population leakage out of one of the aforementioned decay
channels. In the limit that $\Omega(t)$ is small and varies
slowly, all other $c_i(t)$ adiabatically follow $c_{s}(t)$~(see
the Methods section), and one finds
\be
\dot{c}_{e}(t){\approx}-i\Omega(t)c_{s}(t)-\frac{1}{2}\left(\gamma+\frac{4g_1^2}{\kappa_{a}+4g_2^2/\kappa_c}\right)c_{e}.
\ee
Physically, we can identify
$\Gammatotal=\gamma+\frac{4g_1^2}{\kappa_{a}+4g_2^{2}/\kappa_c}$
as the cavity-enhanced total decay rate of $\ket{e}$, where the
first~(second) term corresponds to direct radiative
emission~(emission into mode $a$). Similarly, the denominator
$\kappa_{a}+4g_{2}^{2}/\kappa_c$ corresponds to the new total
decay rate of mode $a$ in the presence of an optical nonlinearity,
as it yields a new channel for photons to effectively ``decay''
out of mode $a$ into $c$ at rate $4g_2^2/\kappa_c$. It is clear
that some optimal value of $g_2$ exists for frequency conversion
to occur. In particular, for no nonlinearity~($g_2=0$) this
probability is non-existent. On the other hand, for
$g_{2}\rightarrow\infty$, one finds $\Gammatotal=\gamma$, which
indicates that the leakage from mode $a$ into $c$ is so strong
that no cavity-enhanced emission occurs. Note that the use of 
time-varying control and pump fields allows for
arbitrary shaping of the outgoing
single-photon wavepacket at frequency $\omega_c$, provided only
that the photon bandwidth is smaller than $\kappa_c$~(physically,
the photon cannot leave faster than the rate determined by the
cavity decay, see Methods). This feature is particularly useful in
two respects. First, in practice $\kappa_c$ can be much larger
than $\gamma$, which enables extremely fast operation times.
Second, pulse shaping is useful for constructing quantum networks,
as it allows one to impedance-match the outgoing photon to other
nodes of the network.

Based on the above arguments, the probability that a single photon
of frequency $\omega_c$ is produced and extracted into the desired
out-coupling waveguide is given by
\be
F=\frac{C_{in}}{1+\phi+C_{in}}\frac{\phi}{1+\phi}\frac{\kappa_{c,ex}}{\kappa_c},
\label{eq:F}
\ee
where $\phi=4g_2^2/(\kappa_{a}\kappa_c)$ characterizes the
branching ratio in mode $a$ of nonlinearity-induced leakage to
inherent losses, and $C_{in}=4g_1^2/\gamma\kappa_a$ is the
inherent cavity cooperativity parameter for mode $a$ in absence of
nonlinearity. The first term on the right denotes the probability for $\ket{e}$
to decay into mode $a$, the second term the probability that a
photon in mode $a$ couples into mode $c$, and the third term the
probability that a photon in mode $c$ out-couples into the desired
channel~(see Methods for an exact calculation).
$\phi$ depends on the pump amplitude $E_b$, with the
optimal value $\phi=\sqrt{1+C_{in}}$ yielding the maximum in $F$.
For large $C_{in}{\gg}1$, the maximum probability is
\be
F{\approx}\left(1-\frac{2}{\sqrt{C_{in}}}\right)\frac{\kappa_{c,ex}}{\kappa_c}.\label{eq:Fmax}
\ee
Considering an emitter placed near the field maximum of mode $a$,
$C_{in}{\sim}\frac{3Q_a}{2\pi^2}\frac{\lambda_a^3}{n_a^{3}V_a}\frac{\gamma_0}{\gamma}$,
where $Q_{a},V_{a}$ are the mode quality factor and volume,
respectively, and $n$ is the index of refraction at frequency
$\omega_a$. The ratio $\gamma/\gamma_0$ is the spontaneous
emission rate into non-cavity modes normalized by the spontaneous
emission rate
$\gamma_{0}{\equiv}n\omega_{a}^{3}|\bfd|^2/(3\pi\epsilon_{0}{\hbar}c^3)$
of an emitter embedded in an isotropic medium of index $n$. This
ratio is expected to be of order $1-10$ for our devices of interest,
and thus the efficiency essentially depends only on $Q_{a}/V_{a}$.

Finally, while we have focused on the case of single-photon
generation here, the reverse process can also be considered, where
a single incoming photon at frequency $\omega_c$ is incident upon
the system, converted into a photon in mode $a$, and coherently
absorbed by an atom with the aid of an impedance-matched pulse
$\Omega(t)$, causing its internal state to flip from $\ket{g}$ to
$\ket{s}$. Generally, by time-reversal
arguments~\cite{gorshkov07}, it can be shown that the probability
$F$ for single-photon storage is the same as that for generation.

\begin{figure}[t]
\centering
\includegraphics[width=15cm]{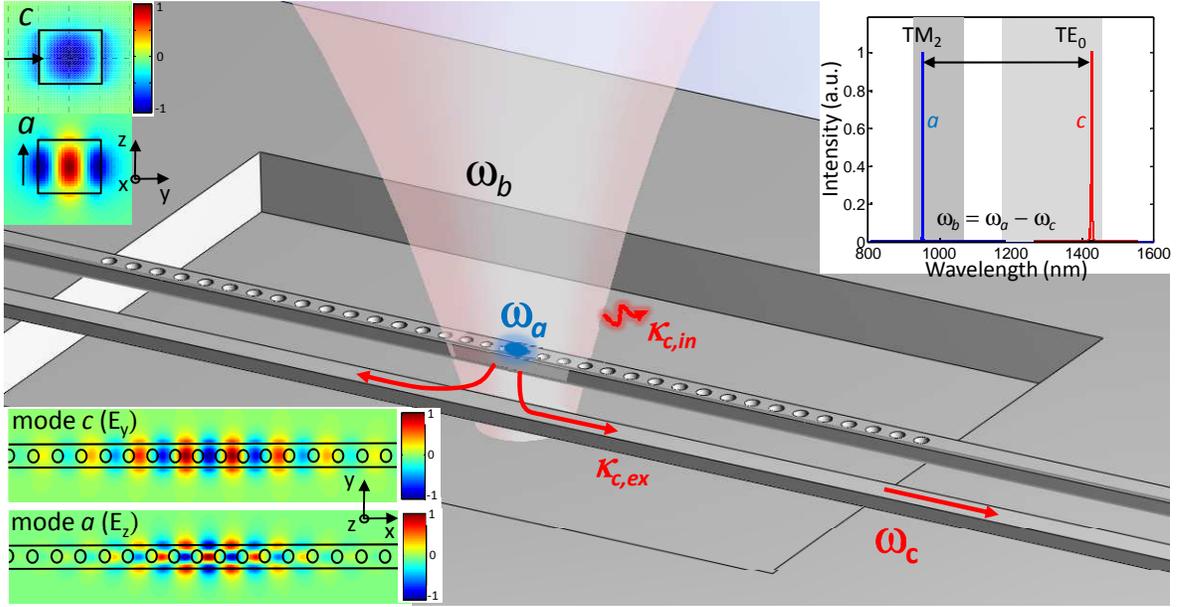}
\caption{\textbf{Cavity mode characteristics} Frequency
conversion platform based on a photonic crystal nanobeam cavity, integrated extraction
waveguide, and off-chip coupling laser ($\omega_b$) tuned to the difference
frequency of the modes.  The cavity is formed by introducing a local perturbation into a 
periodic 1D line of air holes in the free-standing nanobeam.
The desirable ($\kappa_{c,ex}$) and inherent
($\kappa_{c,in}$) loss channels from mode $c$ are shown.
The insets show the schematic cavity spectrum
with photonic stopbands shown in grey, and the dominant field components of the
TE$_0$ ($\omega_c$) and TM$_2$ ($\omega_a$) modes.  The $yz$-plane cross-sections 
of the modes (upper left) show the $E_y$ ($E_z$) component of mode $c$ ($a$)
at the center of the cavity, highlighting the mode overlap and polarizations.  
In the optimized structure (described in Methods), the TE mode
at 1425 nm has $Q = 1.2 \times 10^7$ and $V_n = 0.77$, and the TM
mode at 950 nm has $Q = 7.3 \times 10^4$ and $V_n = 1.45$ ($V_n$ is
the mode volume normalized by $(\lambda/n)^3$). The inherent peak
cooperativities for the modes are $C_{in}^{TE} = 2.4 \times
10^7$ and $C_{in}^{TM} = 3.7 \times 10^4$, which
are well into the strong coupling regime, as given by $C > 1$.
\label{fig:tetm}}
\end{figure}

\section*{Realization in a nonlinear photonic crystal cavity}

In order to implement this frequency conversion scheme in a practical
fashion, there are several constraints on the design of the cavity
modes.  For the nonlinear process to be efficient, mode $a$ must
have a high cooperativity ($Q/V$) to ensure strong
coupling of the emitter (see Fig.~\ref{fig:levels}).
For mode $c$, a high $Q$ factor (small $\kappa_c$) is important to
maximize the nonlinear coupling parameter, $\phi$, and hence
reduce the pump power needed in order to reach the optimum
nonlinear coupling strength, $g_2$.  The cavity should
also be composed of a $\chi^{(2)}$ nonlinear material
that is transparent in the desired frequency range.
Finally, in order for the modes to couple efficiently via
the nonlinear susceptibility of the cavity, they must have a large
spatial overlap and the appropriate vector orientation, as
determined by the elements of the $\chi^{(2)}$ tensor of the
cavity material~(see Eq.~(\ref{eq:g2})).

As a host platform for the nonlinear cavity, the III-V
semiconductors are promising candidates because of their
significant second-order nonlinear susceptibilities and mature
nanofabrication technologies.  However, the symmetry of the III-V
group $\chi^{(2)}$ tensor ($\chi_{ijk}^{(2)}{\neq}0, i\!\neq\! j\!
\neq\! k$) requires that the dominant field components of the
modes be orthogonal in order to maximize the nonlinear coupling.
It further implies that if the classical field which drives the
nonlinear polarization is incident from the normal direction
(\textit{e.g.}, from an off-chip laser), one of the cavity modes
must have a TM polarization.  

We adopt a photonic crystal platform to realize a wavelength-scale nonlinear
cavity that meets these requirements.  Recently, 2D photonic crystal nanocavities 
have shown great promise for strongly coupling an optical mode to a quantum dot 
emitter~\cite{Hennessy07,Englund07}.  In addition, they have been used as platforms
for classical nonlinear optical generation and switching~\cite{McCutcheon_PRB, Tanabe05}.  
The challenge, however, is to design a nonlinear photonic crystal
nanocavity which supports two orthogonal, high cooperativity modes with a
large mode field overlap.  

To enable
a monolithic cavity design which supports {\em both} TE {\em and} TM
modes, we design a photonic crystal ``nanobeam'' cavity -- a free-standing
ridge waveguide patterned with a one-dimensional (1D) lattice of holes --
for which we can control both TE and TM photonic bandstructures.
Recently, there has been much interest in photonic crystal nanobeam
cavities~\cite{Sauvan,Zain_08,McCutcheon_08, Notomi_1D, Deotare09} due to their
exceptional cavity figures of merit ($Q$ and $V$), relative ease of design
and fabrication, and potential as a platform to realize novel optomechanical
effects~\cite{Povinelli,Eichenfeld_08}.  Our frequency conversion scheme can
be realized in a similar structure, as shown in Fig.~\ref{fig:tetm}. We optimize 
two high cooperativity cavity modes by exploiting the different quasi-1D TE and TM
photonic stopbands of the patterned nanobeam 
(shaded regions in the inset of Fig.~\ref{fig:tetm}).
A key design point is that the TE and TM bandstructures can be tuned
somewhat independently by varying the cross-sectional aspect ratio
of the ridge.  For example, in a nanobeam with a square
cross-section, the two stopbands overlap completely. As the
width-to-depth ratio of the waveguide is increased, the effective
index of the TE modes increases relative to the TM modes,
shifting the TE stopband to longer wavelengths.

\subsection*{Example implementations}

As a first example, we design a GaAs photonic crystal nanobeam cavity with modes at 950 nm
and 1425 nm suitable for directly generating single photons at telecom
wavelengths from InAs/GaAs quantum dots. To achieve such a large spectral 
separation, we couple the fundamental TE$_0$ cavity mode to a {\em higher-order} 
TM$_2$ cavity mode (see inset Fig.~\ref{fig:tetm}).  Crucially, the photonic crystal
lattice tapering~\cite{McCutcheon_08,Notomi_1D, Eichenfeld_08} is effective in enhancing the $Q$ 
factors of {\em both TE and TM modes}. Details of the cavity parameters 
and optimization are provided in the Methods section. For this cavity, the 
coupling field ($E_b$) must have a wavelength $\lambda_b = 2.85\;\mu$m in 
order to efficiently drive the difference-frequency process.
GaAs is an attractive
nonlinear cavity material because it has a reasonably large 
$\chi^{(2)}$ strength~\cite{Singh}, a high refractive index, and mature 
microfabrication techniques.  

As evident in Fig.~\ref{fig:tetm}, the overlap of the two
modes changes sign near the edges of the ridge compared to the
middle due to the different symmetries of the TE$_0$ and 
TM$_2$ modes.  However, the induced nonlinear polarization is
dominated by the negatively signed anti-nodes near the middle of
the ridge, and the imperfect overlap in the integral can
be completely compensated for by a stronger pump beam.  Thus, by
selecting a higher order TM$_2$ mode, we have gained a larger
frequency conversion bandwidth at the expense of the somewhat
higher pump power required to overcome the ensuing phase mismatch.
Note that the fundamental TE$_0$-TM$_0$ mode overlap is nearly ideal, 
and would be appropriate for applications requiring relatively small 
frequency shifts.

\begin{figure}[tb]
\centering
$\begin{array}{c@{\hspace{0.5cm}}c}
\includegraphics[width=10cm]{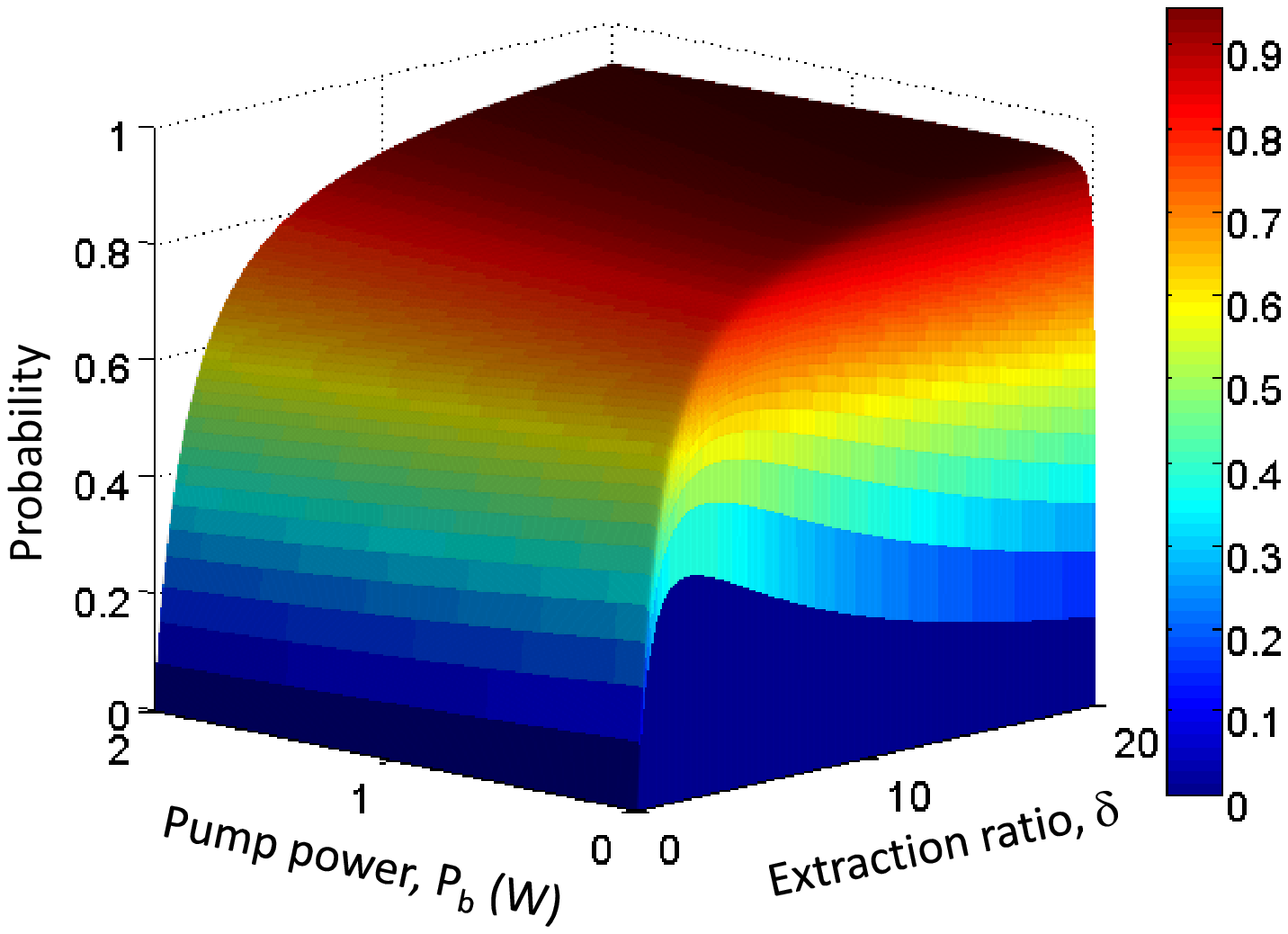} &
\hspace{0cm}\includegraphics[width=8cm]{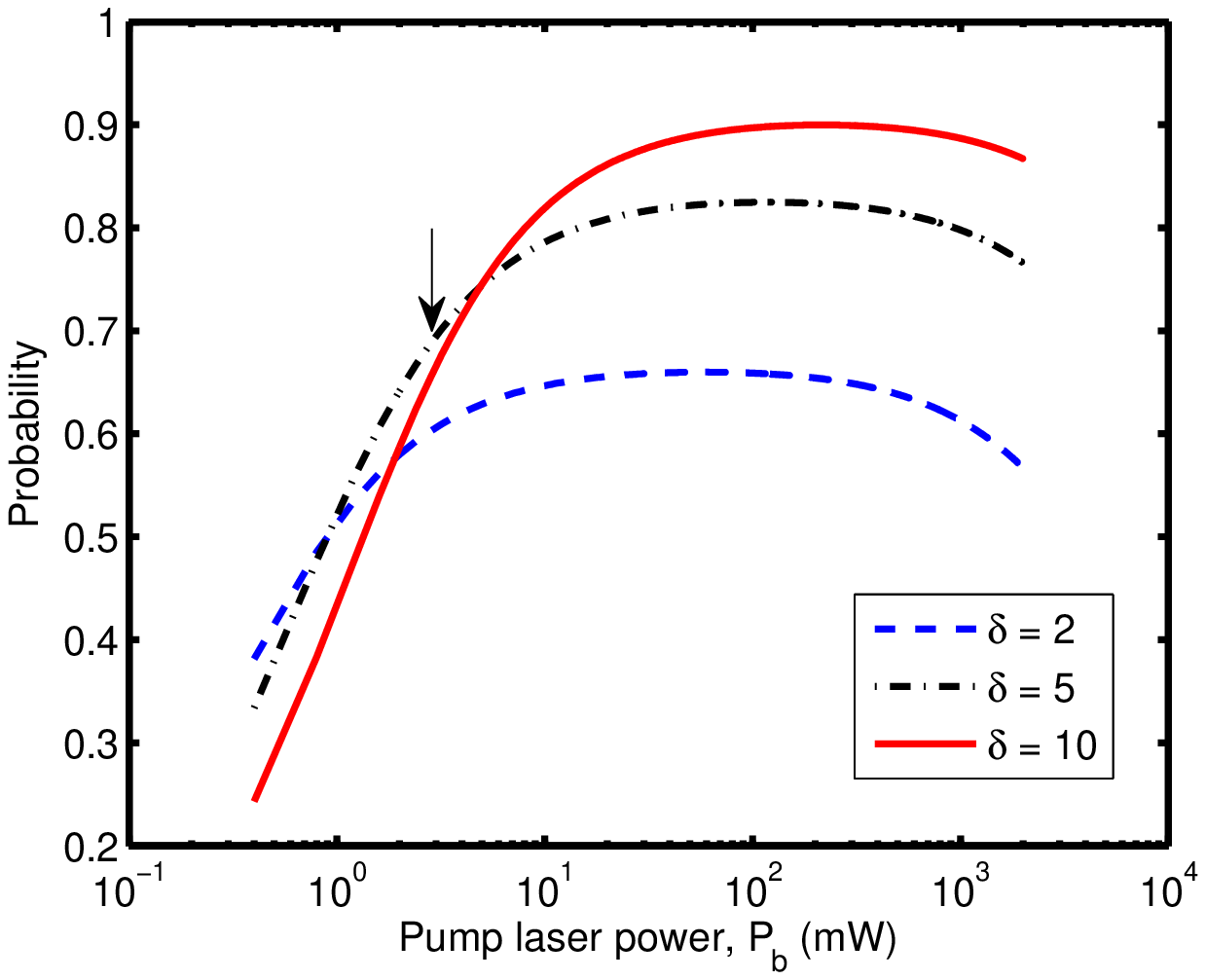} \\
(a) & (b) \\
\end{array}$
\caption{\textbf{Probability of single-photon frequency
conversion} from 950 nm to 1425 nm.  The photon is
coupled into a well-defined output channel at rate
$\kappa_{c,ex}$. Note that the internal probability of conversion in 
the absence of an over-coupled extraction channel is 0.99.
(a) Probability as a function of the pump laser
power, $P_b$, and the extraction ratio, $\delta =
\kappa_{c,ex}/\kappa_{c,in}$.  For a given $\delta$, there is an
optimal operating power, $P_b$, as visible by the sharp contour
ridge at small $P_b$.  (b) Probability as a function of $P_b$ for
different values of $\delta$.  Because of the rapid rise in
probability at low $P_b$, the system does not need to be operated
at the optimum to achieve high conversion probabilities. For
example, for $\delta=10$, a probability of $F=0.7$ can be achieved
with a pump power $P_b$ = 3 mW~(indicated by the arrow).
\label{fig:fidel}}
\end{figure}

We now calculate the probability to convert a single photon from
950 to 1425 nm in our system. The optimized cavity design (see
Methods) simultaneously yields high quality factors and small mode
volumes, which allows for extremely high cooperativities for each
mode ($C_{in} > 10^4$).  From
Eq.~(\ref{eq:Fmax}) we find that this enables an internal
conversion probability of up to $F=0.99$ when waveguide extraction efficiency
is not taken into account.  In practice, to
efficiently out-couple the frequency-converted single photon into
a waveguide, we require the ratio $\delta =
\kappa_{c,ex}/\kappa_{c,in}$ to be large (i.e. overcoupled).
The branching parameter
$\phi$ scales as $P_b/(\kappa_{c,in}(1 +\delta))$, and so to
increase the extraction ratio $\delta$, the pump power~($P_b$)
must also be increased to maintain the optimal $\phi$.
Essentially, achieving good extraction efficiency requires one to
intentionally increase the losses in mode $c$~(via the
out-coupling waveguide), which in turn requires more pump power to
maintain the critical coupling. This relationship is made clear in
Fig.~\ref{fig:fidel}, which plots the probability $F$ as a
function of pump power $P_b$ and extraction ratio $\delta$.  For a
given $\delta$, the power $P_b$ can be chosen to maximize the
probability, reflecting the optimal value of $g_2$ for frequency
conversion.  The probability rises rapidly with $P_b$,  reaching a
maximum at relatively low powers (visible as the sharp ridge in
the contours). Three fixed $\delta$ contours are plotted in
Fig.~\ref{fig:fidel}(b), demonstrating that efficient extraction
of frequency-converted photons can be realized at modest pumping
powers.  For example, for $\delta = 10$, an extraction probability
of 0.7 can be realized with a coupling laser power of 3 mW focused
in a diffraction-limited focal spot. We note that the absorption of
GaAs at 2.85 $\mu$m is negligible, and so there will be no
pump-induced heating.  For this
particular cavity design, the outgoing converted photon can be
shaped to have a bandwidth of up to $\kappa_{c}{\sim}100$~MHz.

By exploiting the scaling properties of Maxwell's equations, it is
straightforward to design a similar cavity in GaP which supports modes at 
637 nm and 950 nm.  GaP is a nonlinear wide bandgap semiconductor which is 
transparent at 637 nm and has recently shown promise for
the microcavity enhancement of diamond NV emission~\cite{Fu,Rivoire}.
Accounting for the exact refractive index dispersion
and $\chi^{(2)}$ strength of GaP, we calculate that the internal frequency conversion 
probability is 0.99, and the extraction probability is 0.7 for 
$< 4$ mW coupling power.  The 637-950 nm span would be sufficient 
to couple any pair of the most relevant quantum 
emitters, namely NV centers in diamond; atoms such as Cs or Rb; and InAs/GaAs 
quantum dots.  Such a cavity could also be
integral to creating a stable, room temperature single-photon
source emitting in the telecom band based on frequency-converted
NV center emission in diamond~\cite{Kurtsiefer}.   Given that 
it may be difficult to span the large spectrum from 637 nm to telecom wavelengths
in a single monolithic design, a 637-950 nm cavity could be the
first stage of a two-step frequency conversion process involving our first
example as the second stage.  More generally, cascading allows our design to be 
extended to cover virtually any frequency span.

\section*{Outlook}

We have shown that high-fidelity, intra-cavity frequency
conversion of single photons from a dipole-like emitter can be
achieved using a two-mode nonlinear cavity pumped by a classical
field. Our general framework is valid for conversion between
arbitrary frequencies, and the efficiency depends only on the
cavity parameter $Q/V$. As realistic implementations, we propose
two different high-cooperativity, double-mode photonic crystal
nanocavities to enable highly efficient coupling from
950 nm -- 1425 nm and 637 nm -- 950 nm, respectively.
Single-photon conversion between these wavelengths would
allow, in the first instance, coupling of InAs/GaAs quantum dot 
emission~\cite{Srinivasan07, Hennessy07, Englund07}
into low-loss optical fibers in the telecom spectrum.  The
second example would facilitate a direct optical connection between two
types of solid-state emitters currently of great interest: a
nitrogen-vacancy center in diamond~\cite{Dutt,Santori,Gaebel06,Hanson} and 
an InAs/GaAs quantum dot.  Integrated together, the two designs could 
allow for cascaded frequency conversion of NV center emission into the 
telecom band.  Further design
improvements should lead to larger frequency spans and also lower
pump power requirements~(\textit{e.g.}, by allowing $\omega_{b}$
to correspond to a third cavity mode).  Although we have emphasized
large frequency shifts in this paper, a smaller shift could be readily
achieved by coupling the TE$_0$ mode with the fundamental TM$_0$ mode,
which has a larger $Q$ factor than the TM$_2$ mode studied here.  
The TE$_0$-TM$_0$ modes have a larger spatial overlap, reducing the 
coupling power required for high probability frequency conversion.

Beyond the aforementioned applications, the techniques described
here can potentially be extended to open up many intriguing
opportunities. For example, the photon emission of a particular
emitter could be shifted into wavelengths where high-efficiency
detectors are available.  It also allows coupling of atomic
emitters such as Cs or Rb with solid-state emitters to
create hybrid atom-photonic chips~\cite{Barclay_06}. In addition,
a number of quantum entanglement schemes for atoms rely on joint
photon emission and subsequent detection to probabilistically
project the atomic system into an entangled
state~\cite{cabrillo99,bose99,duan03}. Such schemes rely on the
indistinguishability of photons emitted from each atom, and
implementing such techniques in nonlinear cavities could allow
entanglement between different types of emitters. In addition, the
protocol described here could be extended for generating
narrow-bandwidth, entangled photon pairs with high efficiency and
repetition rates, which are a valuable resource for applications
such as quantum cryptography~\cite{ekert91}.  Our scheme could
also be applicable in active materials, where laser wavelengths
could be converted from easily accessible regions like 1500 nm to
the mid-infrared range.  Finally, it would be intriguing to
combine these ideas with cavities that exhibit opto-mechanical
coupling~\cite{Eichenfeld_08}, which would potentially allow the
photonic frequencies to be dynamically and rapidly tuned.

\section*{Methods}
\subsection*{Derivation of nonlinear conversion efficiency}

The state amplitudes of the wave-function given in
Eq.~(\ref{eq:wavefn}) evolve under the interaction Hamiltonian
$H_I$ of Eq.~(\ref{eq:HI}) through the following equations,
\bea \dot{c}_{s} & = & -i\Omega(t)c_{e},\nonumber \\ \dot{c}_{e} &
= & -i\Omega(t)c_{s}-ig_{1}c_{a}-(\gamma/2)c_{e},\nonumber \\
\dot{c}_{a} & = & -ig_{1}c_{e}-ig_{2}c_{c}-(\kappa_a/2)c_{a},\nonumber \\
\dot{c}_{c} & = &
-ig_{2}c_{a}-(\kappa_{c}/2)c_{c}.\label{eq:diffeq} \eea
These equations describe both coherent evolution~(terms
proportional to $\Omega(t),g_{1,2}$) and population loss in the
system~(terms proportional to $\gamma,\kappa_{a,c}$). The
population loss in the system can be connected to direct radiative
emission of the excited state $\ket{e}$~(at a rate
$\gamma|c_{e}|^2$), radiation leakage and absorption losses of
mode $a$~($\kappa_{a}|c_a|^2$), and absorption and leakage out of
mode $c$~($\kappa_{c}|c_c|^2$, of which $\kappa_{c,ex}|c_c|^2$ is
successfully out-coupled to a waveguide). In general the
efficiency of extracting a single photon of frequency $\omega_{c}$
out into the waveguide is thus
\be
F=\frac{\int_{0}^{\infty}dt\,\kappa_{c,ex}|c_{c}(t)|^2}{\int_{0}^{\infty}dt\,\kappa_{c}|c_{c}(t)|^2+\kappa_{a}|c_a(t)|^2+\gamma|c_{e}(t)|^2}.\label{eq:Fgeneral}
\ee
For arbitrary $\Omega(t)$, Eqs.~(\ref{eq:diffeq})
and~(\ref{eq:Fgeneral}) can be evaluated numerically. However, in
certain limits one can find approximate solutions.  In particular,
when $\Omega(t)$ and its rate of change are small compared to the
natural oscillation and decay rates of the system, the state
amplitudes $c_{a,c,e}$ will follow the instantaneous value of
$c_{s}(t)$. Formally, we can adiabatically eliminate these states,
setting $\dot{c}_{i}=0$ for $i=a,c,e$. Then, one finds
\be
\dot{c}_{s}(t)=-\frac{2\Omega(t)^2}{\Gammatotal}c_{s}(t),\label{eq:csdot}
\ee
while the other $c_i{\propto}c_{s}(t)$, with the proportionality
coefficients being functions of
$g_{1},g_{2},\kappa_{a},\kappa_{c},\Omega(t)$. The resulting
substitution of the solutions of $c_{i}(t)$ into
Eq.~(\ref{eq:Fgeneral}) allows great simplification because the
integrands now become time-independent, and after some
simplification yields Eq.~(\ref{eq:F}). Self-consistency of the
adiabatic elimination solution requires that the the effective
rate of population loss ${\sim}4\Omega(t)^2/\Gammatotal$ predicted
from state $\ket{s}$ does not exceed the rate $\kappa_c$ that a
photon can leak out through the cavity mode $c$.

In the effective wave-function approach used here, the population
leakage out of mode $c$ can also be explicitly related to the
shape of the outgoing single-photon wavepacket. For instance, we
can model the linear coupling of cavity mode $c$ to photons
propagating in a single direction in a waveguide with the
following Hamiltonian~(in a rotating frame),
\be
H_{w}=\int\,dk\,{\hbar}v(k-\omega_{c}/v)\opdagger{a}{k}\op{a}{k}-\hbar{g_w}\int\,dk\,\left(\opdagger{a}{c}\op{a}{k}e^{ikz_c}+h.c.\right).
\ee
Here $k$ denotes the set of wavevectors of the continuum of
waveguide modes, $v$ is the velocity of waveguide fields, $g_w$ is
the coupling strength between cavity and waveguide modes, and
$z_c$ denotes the position along the waveguide where the cavity is
coupled to it~(for simplicity we set $z_c=0$ from this point on).
Since we are now explicitly accounting for the waveguide degrees
of freedom, we add a term $\int\,dk\,c_{k}(t)\ket{1_k}$ to the
effective wave-function of the system. The equations of motion of
the total system are identical to Eq.~(\ref{eq:diffeq}), except
that
\bea \dot{c}_{c} & = &
-ig_{2}c_{a}-(\kappa_{c,in}/2)c_{c}+ig_{w}\int\,dk\,c_{k}, \\
\dot{c}_{k} & = & -iv(\delta{k})c_{k}+ig_{w}c_{c}, \eea
where ${\delta}k=k-\omega_{c}/v$. Compared to
Eq.~(\ref{eq:diffeq}), we have now included the coupling of mode
$c$ to the waveguide, and accordingly have replaced
$\kappa_{c}{\rightarrow}\kappa_{c,in}$ in the equation for
$\dot{c}_{c}$ since the leakage into the waveguide should be
accounted for by the new coupling terms. The equation for
$\dot{c}_{k}$ can be formally integrated; assuming that the
waveguide initially is unoccupied, $c_{k}(0)=0$, one has
\be
c_{k}(t)=ig_{w}\int_{0}^{t}dt'\,c_{c}(t')e^{-ic{\delta}k(t-t')}.\label{eq:ck}
\ee
Substituting this into the equation for $\dot{c}_{c}$ and
performing the Wigner-Weisskopf approximation~\cite{scully97}, one
recovers the expression for $\dot{c}_{c}$ in Eq.~(\ref{eq:diffeq})
by identifying $\kappa_{c,ex}=2{\pi}g_{w}^2/v$. The one-photon
wave-function~\cite{scully97} is given by
$\psi_{w}(z,t)=\bra{vac}\hat{E}_{w}(z,t)\ket{\psi(0)}=(\sqrt{2\pi}ig_{w}/v)\Theta(z)c_{c}(t-z/v)$,
where $\Theta(z)$ is the step function. The wave-function shape is
thus directly proportional to $c_{c}(t)$. Under adiabatic
elimination,
\be
\psi_{w}(z,t)=\frac{\sqrt{2{\pi}}ig_{w}}{v}\Theta(z)\frac{8ig_{1}g_{2}}{\Gammatotal(\kappa_{a}\kappa_{c}+4g_2^2)}\Omega(t-z/v)c_{s}(t-z/v),\label{eq:psiw}
\ee
and thus for a desired~(and properly normalized) pulse shape
$\psi_{w}$ one needs only to solve Eqs.~(\ref{eq:psiw})
and~(\ref{eq:csdot}) to obtain the corresponding external field
$\Omega(t)$. It is straightforward to show that the normalization
is given by $\int\,dz\,|\psi_{w}(z,t{\rightarrow}\infty)|^2=F$
provided that $c_{s}(\infty){\rightarrow}0$. This normalization
reflects the probability that a single photon ends up in the
waveguide.

\subsection*{Nonlinear cavity design}

The nanobeam cavities are formed by a 4-period taper in the size
and spacing of the holes in the uniform photonic ``mirror'' on
both sides of the cavity center in order to introduce a localized
potential for the TE and TM modes.  The 950-1425 nm cavity nanobeam
has width $w$ = 420 nm and depth $d$ = 307.5 nm, and the hole
spacing tapers from $a_0$ = 360 nm in the mirror to $a_c$ = 337 nm
in the center.  The holes were made elliptical to give an
additional design parameter to separately optimize the TE and TM
mode $Q$ factors.  The elliptical hole semi-axes are 84 nm and 108
nm in the mirror section, and the hole size-to-spacing ratio is
held constant through the taper section. This design yields cavity
parameters of $Q = 1.2 \times 10^7$ and $V_n = 0.77$ for the TE
mode, and $Q = 7.3 \times 10^4$ and $V_n = 1.45$ for the TM
mode~($V_n$ is the mode volume normalized by $(\lambda/n)^3$). The
factor $\gamma/\gamma_0$ = 0.10 (0.20) for the TE (TM) mode is
determined by simulating the total power emitted by a non-resonant
dipole source in the cavity center.  We have accounted for the
index dispersion of our candidate material, GaAs, for which $n$(1425
nm) = 3.38 and $n$(950 nm) = 3.54~\cite{Palik}. 

The nonlinear parameter $g_2$ is determined by calculating the
volume integral of Eq.~(\ref{eq:g2}) using the exact mode fields,
$E_a$ and $E_c$, extracted from our 3D-FDTD calculation. Because
the mode fields are oriented along the $y$ and $z$-axes,
respectively, as defined in Fig.~\ref{fig:tetm}(c), the classical
field which drives the differency frequency generation, $E_b$,
must be polarized along $x$.  This field has a frequency $\omega_b
= \omega_a - \omega_c$, which corresponds to a wavelength
$\lambda_b = 2.85\;\mu$m.  The relevant nonlinear
susceptibility tensor elements are $\chi_{xyz}^{(2)}(GaAs) = 2 d_{14} =
550$ pm/V and $\chi_{xyz}^{(2)}(GaP) = 320$ pm/V~\cite{Singh, Shoji}.

We assume the
classical field is constant over the spatial extent of the cavity
modes, which allows $E_{b,x}$ to be taken in front of the integral
for $g_2$, giving
\begin{equation}
g_2 = - \frac{\epsilon_0 E_{b,x}}{\hbar} \int d{\bf r}
\chi^{(2)}_{xyz} E^*_{a,y} E_{c,z}. \label{eq:beta2}
\end{equation}
To justify this assumption, we simulated a Gaussian beam
with $\lambda_b = 2.85\; \mu$m that is focused by a lens with a modest
numerical aperture (NA) of 0.5 onto a ridge waveguide, and found that
the average field amplitude is approximately
uniform over the linear extent of our cavity modes (approx. $x$ = -1
$\mu$m to +1 $\mu$m).  In the $g_2$ calculation, the magnitude of
$E_{b,x}$ for a given beam power, $P_b$, is then determined from the
relation $P_b = \epsilon_0 c \pi r^2 E_{b,x}^2/4$, where $r$ is the focal 
spot radius.

\section*{Acknowledgements}

MWM would like to thank NSERC (Canada) for its support, and DEC acknowledges support 
from the Caltech CPI.  The authors also gratefully acknowledge useful discussions with 
Jelena Vu\v{c}kovi\'{c}.


\end{document}